\documentclass[12pt]{article}
\usepackage{latexsym}
\usepackage{amsmath}
\usepackage{amsfonts}
\usepackage{amssymb}
\usepackage{amscd}
\usepackage{bbm}
\usepackage{fancybox}
\usepackage{cite}
\usepackage{amsmath,amsfonts,amsbsy}
\pdfoutput=1
\usepackage{graphicx}
\usepackage{epsfig}
\usepackage{hyperref}

\usepackage[utf8x]{inputenc}
\usepackage{bbold}

\usepackage{float}

\usepackage{multirow}                     
\usepackage{float}                          
\usepackage{lscape}                         
\usepackage{bm}

\newcommand{\cG}{{\cal G}}
\newcommand{\cJ}{{\cal J}}

\newcommand{\cL}{{\cal L}}
\newcommand{\cM}{{\cal M}}
\newcommand{\cN}{{\cal N}}

\newcommand{\cQ}{{\cal Q}}
\newcommand{\cR}{{\cal R}}

\newcommand{\bb}{\bar\beta}

\newcommand{\beq}{\begin{equation}}
\newcommand{\eeq}{\end{equation}}
\newcommand{\bi}{\begin{itemize}}
\newcommand{\ei}{\end{itemize}}
\newcommand{\bt}{\begin{tabular}}
\newcommand{\et}{\end{tabular}}
\newcommand{\bc}{\begin{center}}
\newcommand{\ec}{\end{center}}

\newcommand{\Str}{\text{STr}}
\newcommand{\extd}{\text{d}}

\def\one{{\hbox{ 1\kern-.8mm l}}}
\newcommand{\Dslash}{\not{\hbox{\kern-4pt $D$}}}
\newcommand{\pdslash}{\not{\hbox{\kern-2pt $\partial$}}}

\newcommand{\be}{\begin{equation}}
\newcommand{\ee}{\end{equation}}
\newcommand{\bea}{\begin{eqnarray}}
\newcommand{\eea}{\end{eqnarray}}
\newcommand{\ba}{\begin{array}}
\newcommand{\ea}{\end{array}}

\def\bbox{{\,\lower0.9pt\vbox{\hrule \hbox{\vrule height 0.2 cm
\hskip 0.2 cm \vrule height 0.2 cm}\hrule}\,}}
\newcommand{\dsl}{\pa \kern-0.5em /}

\newcommand{\vp}{\varphi}

\font\mybb=msbm10 at 12pt
\def\bb#1{\hbox{\mybb#1}}
\def\bZ {\bb{Z}}
\def\bR {\bb{R}}

\begin{document}

\begin{titlepage}
\begin{center}

\hfill TUW-16-21

\vskip 1.5cm

{\Large \bf Super-BMS$_3$ algebras from $\mathcal{N}=2$ flat supergravities}

\vskip 1cm

{\bf Ivano Lodato$^1$ and Wout Merbis$^2$}

\vskip 25pt

{\em $^1$ \hskip -.1truecm Indian Institute of Science Education and Research,\\ Homi Bhabha Road, Pashan, Pune 411 008, India \\ \vskip 5pt }

{email: {\tt ivano@iiserpune.ac.in}} \\
\vskip 10pt

{\em $^2$ \hskip -.1truecm  Institute for Theoretical Physics, Vienna University of Technology, \\
Wiedner Hauptstrasse 8-10/136, A-1040 Vienna, Austria\vskip 5pt }
{email: {\tt merbis@hep.itp.tuwien.ac.at}} \\
\vskip 10pt

\end{center}

\vskip 1.5cm

\begin{center} {\bf ABSTRACT}\\[3ex]
\end{center}

\noindent
We consider two possible flat space limits of three dimensional $\mathcal{N} = (1,1)$ AdS supergravity. They differ by how the supercharges are scaled with the AdS radius $\ell$: the first limit (democratic) leads to the usual super-Poincar\'e theory,  while a novel `twisted' theory of supergravity stems from the second (despotic) limit. We then propose boundary conditions such that the asymptotic symmetry algebras at null infinity correspond to supersymmetric extensions of the BMS algebras previously derived in connection to non- and ultra-relativistic limits of the $\mathcal{N}=(1,1)$ Virasoro algebra in two dimensions. Finally, we study the supersymmetric energy bounds and find the explicit form of the asymptotic and global Killing spinors of supersymmetric solutions in both flat space supergravity theories.

\end{titlepage}

\newpage

\section{Introduction}
The choice of boundary conditions for a certain field configuration in three dimensional gravity not only limits the allowed background solutions, but it is also intimately connected to its asymptotic symmetries. For instance, in the seminal work \cite{Brown:1986nw} it was shown that imposing suitable boundary conditions on fields in an asymptotically AdS$_3$ space enhances the symmetries at spatial infinity from the SO(2,2) isometries of AdS space to the infinite dimensional conformal algebra in two dimensions. Furthermore, new excited solutions arise with the same asymptotics but quite different behaviour in the bulk \cite{BTZ,NavarroSalas:1999sr,Nakatsu:1999wt}.  Recently, a great deal of interest has been devoted to a deeper understanding of certain boundary-condition-preserving symmetries, including the symmetries of the conformal boundary of Minkowski space: the Bondi-Metzner-Sachs (BMS) symmetries \cite{Bondi:1962,Sachs:1962}. Such symmetries, which act at null infinity, are infinite dimensional and include the Poincar\'e algebra as a subalgebra.
Although, in their wake, BMS symmetries arose as symmetries preserving asymptotic flatness, more recently they have been studied in connection to locally non-flat spaces. In particular, they have been analyzed in relation to other null-surfaces, such as black hole horizons, either directly \cite{Hawking:2016msc,Penna:2015gza} or indirectly, as composite algebras which may be constructed form a `near-horizon' symmetry algebra \cite{Donnay:2015abr,Afshar:2016wfy,MoreHair}. 

In three spacetime dimensions, BMS$_3$ symmetries of asymptotically flat solutions at null infinity \cite{Ashtekar:1996cd,Barnich:2006av} and the conformal group of asymptotic symmetries of AdS$_3$ spaces at spatial infinity have been connected to each other via a limiting procedure, where the AdS radius is sent to infinity \cite{Barnich:2012aw}. Algebraically, this corresponds to an \.In\"on\"u-Wigner contraction of the Virasoro algebra to the BMS$_3$ algebra, but it also has important physical consequences for the solutions of the theory. For example, an appropriate infinite radius limit of a BTZ black hole corresponds to `flat space cosmologies' \cite{Cornalba:2002nv,Cornalba:2002fi,Barnich:2012xq,Bagchi:2012xr}. Of course, it is a relevant and interesting question to ask whether the set of boundary conditions can be enlarged to accommodate for unbroken supercharges, hence obtaining supersymmetric generalizations of BMS$_3$. Although supersymmetric extensions of BMS have been known since the work of \cite{Grisaru:1977kk,Awada:1985by}, the precise boundary conditions for the gauge field describing the theory to obtain the $\cN=1$ super BMS$_3$ algebra asymptotically were first presented quite recently in \cite{Barnich:2014cwa}. 

In this paper we aim to extend the results of \cite{Barnich:2014cwa} to include the first extended supersymmetric, $\cN=2$ case. Our starting point in section \ref{sec:sugra} will be the construction of a $\cN=2$ theory of supergravity following from a flat space limit of the $\cN = (1,1)$ AdS$_3$ supergravity of \cite{Achucarro:1987vz}. We then impose suitable boundary conditions on the gauge fields and compute the asymptotic symmetry algebra of boundary condition preserving transformations. 

At first sight this may appear to be a rather straightforward generalization of known results; however, as it was shown in \cite{deAzcarraga:2009ch,Sakaguchi:2009de,Mandal:2010gx,projektarbeit,Casali:2016atr,Bagchi:2016yyf,Banerjee:2016nio} at the level of the algebra, the limiting procedure is not unique. In this paper, we construct two different $\cN=2$ relativistic flat supergravity theories as a limit of $\cN = (1,1)$ AdS$_3$ supergravity\footnote{It is also possible to start from the (2,0) (or equivalently (0,2)) AdS$_3$ supergravity algebra, which contains R-symmetry, and obtain new $\cN=2$ theories. These cases will not be treated in the following.}. One is obtained by scaling both left and right moving fermionic generators in the same way, which we will call the \emph{democratic} limit. Another flat space limit may be defined by combining the fermions into two new generators which scale unevenly as the AdS radius goes to infinity. We will call the latter limit \emph{despotic}. This leads to a novel $\cN=2$ `twisted' theory of supergravity, which, to the best of our knowledge, has not been analyzed before.

After constructing the two flat space supergravity theories, we study the boundary conditions at null infinity in section \ref{sec:bcs}, leading to two different $\cN=2$ super BMS$_3$ algebras. The two asymptotic symmetry algebras are isomorphic to the two super-Galilean conformal algebras discussed in \cite{Bagchi:2016yyf} in the context of the tensionless superstrings. The democratic symmetry algebra was found also in \cite{Casali:2016atr,Banerjee:2016nio}, while the despotic algebra was previously obtained in \cite{deAzcarraga:2009ch,Sakaguchi:2009de,Mandal:2010gx} as a Galilean limit of the $\cN = (1,1)$ superconformal algebra. As we  will show below, however there are subleties associated to the non- and ultra-relativisitic limits of the super-Virasoro algebra that lead to the despotic algebra. To avoid any confusion, in the following we will refer to these algebras as $\cN=2$ super-BMS$_3$ algebras, stressing their relativistic origin.
Finally, in section \ref{sec:Killspin} we will compute the energy bounds imposed by supersymmetry, comment on the (supersymmetric) ground states solutions of both theories and solve the asymptotic and exact Killing spinor equations. For simplicity of comparison, each section treats both limits in parallel.

\section{Two flat space limits}\label{sec:sugra}
Our starting point will be $\cN = (1,1)$ AdS$_3$ supergravity \cite{Achucarro:1987vz,Henneaux:1999ib}, defined as the Chern-Simons theory with action
\begin{equation}\label{CSaction}
I[A] = \frac{k}{4\pi} \int \Str \left(A \wedge  \extd A + \frac23 A \wedge A \wedge A \right)\,.
\end{equation}
The one-form gauge connection $A$ takes values in $Osp(1|2,\bR)_+\otimes Osp(1|2,\bR)_-$ 
\begin{equation}
A = A^{+\,a} J_a^{+} + \psi^{+\,\alpha} Q^+_{\alpha} + A^{-\,a} J_a^{-} + \psi^{-\,\alpha} Q^-_{\alpha}\,,
\end{equation}
where the generators $J_a^{\pm}$ and $Q^{\pm}_{\alpha}$ span the $Osp(1|2,\bR)$ algebra
\begin{subequations}\label{Osp12}
\begin{align}
[J^{\pm}_a, J^{\pm}_b] & = \epsilon_{abc}J^{\pm\,c} \,,  \\
[J^{\pm}_a, Q^{\pm}_{\alpha}] & = \frac12 (\Gamma_a)^{\beta}{}_{\alpha} Q_{\beta}^{\pm}\,, \\
\{Q^{\pm}_{\alpha}, Q^{\pm}_{\beta} \} & =  - \frac12 (C\Gamma^a)_{\alpha\beta} J^{\pm}_a\,.
\end{align}
\end{subequations} 
Here $a,b,c = 0,1,2$ and $\alpha,\beta =\pm 1/2$ and our conventions for the gamma matrices $\Gamma_a$ are listed in appendix \ref{sec:conventions}. All commutators which mix the chiral sectors vanish and the trace in the action \eqref{CSaction} is normalized as
\begin{equation}\label{StrAdS}
\Str(J^{\pm}_a J^{\pm}_b) = \pm \frac12 \eta_{ab} \,, \qquad \Str(Q^{\pm}_{\alpha} Q^{\pm}_{\beta}) = \pm \frac12 C_{\alpha\beta}\,.
\end{equation}
We are interested in taking a flat limit of this action, so we should introduce the AdS length $\ell$ explicitly in the action and make a suitable redefinition of the fields such that the limit is finite. To make the connection with the $\cN =(1,1)$ AdS$_3$ supergravity action explicit, we first write
\begin{align}\label{scale}
A^{\pm\,a} = \omega^a \pm \frac{1}{\ell} e^a\,, \qquad k = \frac{\ell}{4G}\,.
\end{align}
Here $\omega^a$ is the dualized spin connection and $e^a$ is the dreibein. The action \eqref{CSaction} reduces to
\begin{align}
I = \frac{1}{16\pi G_N} \int \bigg\{ 2\left(e_a  R^a + \frac{1}{\ell^2} e \right) & - \frac{\ell}{2}\left( \bar{\psi}^+  D \psi^+ + \frac{1}{2\ell} \bar{\psi}^+  e^a \Gamma_a  \psi^+ \right)  \\
&  + \frac{\ell}{2}\left( \bar{\psi}^-  D \psi^- - \frac{1}{2\ell} \bar{\psi}^-  e^a \Gamma_a  \psi^- \right) \bigg\}\,. \nonumber
\end{align}
Here and in the rest of the paper, we have suppressed wedge products and define $D\psi = \extd\psi + \frac12 \omega^a \Gamma_a  \psi$ and $R^a = \extd\omega^a + \tfrac12\,\epsilon^a{}_{bc}\omega^b  \omega^c$. We denote the determinant of the dreibein by $e$.  

For a finite $\ell \to \infty$ limit, the two gravitini $\psi^{\pm}$ need to be rescaled. The conventional way to do this is to scale both of them democratically by a factor of $1/\sqrt{\ell}$. 
\begin{equation}\label{democraticpsi}
\psi^{+} = \sqrt{\frac{2}{\ell}} \psi \,, \qquad \psi^- = \sqrt{-\frac{2}{\ell}} \eta\,.
\end{equation}
Here the factor of $i$ in the redefinition of $\psi^-$ is to ensure the same sign for the kinetic terms of the fermions. The resulting flat space limit is obtained by taking  $\ell \to \infty$ and the action becomes
\begin{equation}\label{Idem}
I_{\rm dem} = \frac{1}{16\pi G_N} \int \left\{ 2 e_a  R^a  - \bar{\psi}  D \psi - \bar{\eta}  D \eta  \right\}\,. 
\end{equation}
We will refer to this limit as the democratic limit.

Interestingly, the democratic scaling \eqref{democraticpsi} is not the only way to obtain a finite action when sending $\ell \to \infty$. Another possibility is to redefine the fermionic fields in a manner similar to the bosonic part of the gauge connection $A$:
\begin{equation}\label{despoticpsi}
\psi^{\pm} = \psi \pm \frac{1}{\ell} \eta\,.
\end{equation}
We end up with a well defined action in the limit $\ell \to \infty$, but different from \eqref{Idem}. Now we obtain
\begin{equation}\label{Idesp}
I_{\rm desp} = \frac{1}{16\pi G_N} \int \left\{ 2 e_a  R^a  - \bar{\psi}  D \eta - \bar{\eta}  D \psi - \frac12 \bar{\psi}  e^a \Gamma_a  \psi \right\}\,. 
\end{equation}
We shall call this limit the despotic limit.

In the next two subsections, we will analyze the two limits as two different contractions of the $Osp(1|2,\bR)_+\otimes Osp(1|2,\bR)_-$ algebra and perform some basic checks of the resulting theories. Readers familiar with the conventional $\cN=2$ flat space supergravity \eqref{Idem} may want to skip straight to section \ref{sec:desplimit} where we treat the despotic limit.

\subsection{Democratic flat space limit}
The democratic action above can similarly be obtained as a Chern-Simons theory of a contraction of the $Osp(1|2,\bR)_+\otimes Osp(1|2,\bR)_-$ algebra by defining new generators
\begin{align}\label{demscaling}
P_a = \frac{1}{\ell}(J^+_a - J^-_a)\,, && J_a = J^+_a + J^-_a\,, && \cQ^{\pm}_\alpha = \sqrt{\pm\frac{2}{\ell}} Q^{\pm}_\alpha\,.
\end{align}
Using the commutation relations \eqref{Osp12} and taking $\ell \to \infty$ we find that these generators span the $\cN=2$ super Poincar\'e algebra, albeit with no R-symmetry.
\begin{subequations}
\label{democratic2}
\begin{align}
[J_a,J_b] & = \epsilon_{abc}J^c\,, &&& [J_a, P_b] & = \epsilon_{abc}P^c\,, &&& [P_a,P_b] & = 0\,, \\
[J_a,\cQ^{\pm}_{\alpha}] & = \frac12(\Gamma_a)^{\beta}{}_{\alpha} \cQ^{\pm}_{\beta}\,, &&& [P_a,\cQ^{\pm}_{\alpha}] & = 0\,, \\
\{\cQ^{\pm}_{\alpha}, \cQ^{\pm}_{\beta} \} & = - \frac12 (C\Gamma^a)_{\alpha\beta}P_a\,, &&& \{\cQ^{\pm}_{\alpha}, \cQ^{\mp}_{\beta} \} & = 0\,.\label{QQPdem}
\end{align}
\end{subequations}
The democratic action is nothing else than the Chern-Simons action \eqref{CSaction}, now with level $k = 1/(4G_N)$,\footnote{The factor of $\ell$ in the Chern-Simons level \eqref{scale} is cancelled by a factor of  $1/\ell$ coming from the trace \eqref{StrAdS} using the relations \eqref{demscaling}} and the connection $A$ is parametrized as:
\begin{equation}\label{Adem}
A= e^a P_a + \omega^a J_a + \psi^{\alpha}\cQ^+_{\alpha} + \eta^{\alpha} \cQ^-_{\alpha}\,.
\end{equation}
The supertrace now has non-zero components
\begin{equation}\label{Strdemocratic}
\Str(J_aP_b) = \eta_{ab} \,,\qquad \Str(\cQ^{\pm}_{\alpha}\cQ^{\pm}_{\beta}) = C_{\alpha\beta} \,.
\end{equation}
The action \eqref{Idem} is invariant off-shell under the supersymmetry transformation laws\\ $\delta A = \extd\lambda + [A, \lambda]$ with $\lambda = \epsilon^{\alpha} \cQ^+_{\alpha} + \zeta^{\alpha} \cQ^-_{\alpha}$. In terms of the fields these transformations read:
\begin{subequations}
\begin{align}
\delta e_{\mu}{}^a & = - \frac12 \left(\bar{\epsilon} \Gamma^a \psi_{\mu} + \bar{\zeta} \Gamma^a \eta_{\mu} \right) \, && \delta \omega_{\mu}{}^a  = 0 \,, \\
\delta \psi_{\mu} & = D_{\mu}\epsilon && \delta \eta_{\mu}  = D_{\mu}\zeta\,.
\end{align}
\end{subequations}
The commutator of two supersymmetry transformations closes on-shell into a general coordinate transformation, a Lorentz transformation (with dualized parameter $\lambda^a = \epsilon^{abc}\Lambda_{bc}$) and a supersymmetry transformation with parameters: 
\begin{align}
[\delta_{\rm susy}(\epsilon_1,\zeta_1),\delta_{\rm susy}(\epsilon_2,\zeta_2)] = & \; \delta_{\rm g.c.}(\xi^{\nu} = -\tfrac12 (\bar\epsilon_2 \Gamma^{\nu}\epsilon_1 + \bar{\zeta}_2 \Gamma^\nu \zeta_1)) \\ \nonumber 
& + \delta_{\rm  Lor} (\lambda^a = - \xi^{\nu}\omega_{\nu}{}^a) + \delta_{\rm susy}(\epsilon = - \xi^{\nu}\psi_{\nu}, \zeta = - \xi^{\nu}\eta_{\nu})\,.
\end{align}
The closure of the supersymmetry algebra is on-shell, so in verifying the above we used the field equations
\begin{equation}
R^a = 0\,, \qquad T^a =-\tfrac14\, \left( \bar{\psi}  \Gamma^a \psi + \bar{\eta}\Gamma^a \eta \right) \,, \qquad D\psi= 0\,, \qquad D\eta = 0\,.
\end{equation}
where the torsion tensor is defined as $T^a = \extd e^a + \epsilon^{abc}\omega_b e_c$.

\subsection{Despotic flat space limit}\label{sec:desplimit}

As in the democratic limit, also the despotic action \eqref{Idesp} can be obtained from a contraction of the $Osp(1|2,\bR)_+\otimes Osp(1|2,\bR)_-$ algebra. The field transformations \eqref{despoticpsi} induce the redefinitions
\begin{align}\label{despscaling}
P_a = \frac{1}{\ell}(J^+_a - J^-_a)\,, && J_a = J^+_a + J^-_a\,, &&  \cR_\alpha = \frac{1}{\ell}( Q^{+}_\alpha - Q^{-}_\alpha)\,,
 && \cG_\alpha =  Q^{+}_\alpha + Q^{-}_\alpha\,.
\end{align}
These generators span an algebra which in the $\ell \to \infty$ limit reads
\begin{subequations}
\label{despoticalg}
\begin{align}
[J_a,J_b] & = \epsilon_{abc}J^c\,, &&& [J_a, P_b] & = \epsilon_{abc}P^c\,, \\
[J_a,\cG_{\alpha}] & = \frac12(\Gamma_a)^{\beta}{}_{\alpha} \cG_{\beta}\,, &&& [P_a,\cG_{\alpha}] & = \frac12(\Gamma_a)^{\beta}{}_{\alpha} \cR_{\beta} = [J_a,\cR_{\alpha}]  \\
\{\cG_{\alpha}, \cG_{\beta} \} & = - \frac12 (C\Gamma^a)_{\alpha\beta}J_a\,, &&& \{\cG_{\alpha}, \cR_{\beta} \} & = - \frac12 (C\Gamma^a)_{\alpha\beta}P_a\,.\label{QQPdes}
\end{align}
\end{subequations}
Note that this algebra still has an $Osp(1|2,\bR)$ subalgebra spanned by $J_a$ and $\cG_{\alpha}$. This explains the mass term for $\psi$ in the action. Due to this term, the curvature equation obtains a contribution from the gravitini
\begin{equation}
R^a =-\tfrac14\, \bar{\psi} \Gamma^a \psi \,, 
\end{equation}
This mass term would allow for the possibility of asymptotically non-flat solutions, if the gravitini $\psi$ takes on a non-zero expectation value. Such a solution would break supersymmetry and one would still need to check the consistency with the other field equations
\begin{equation}
T^a =-\tfrac14  \bar{\psi}\Gamma^a \eta \,, \qquad D\psi= 0\,, \qquad D\eta + \frac12 e^a \Gamma_a \psi = 0\,.
\end{equation}
Despite the possibility to have non-flat solutions, the supersymmetric ground state of this theory is Minkowski space and hence we will call this a flat space supergravity. 

The action \eqref{Idesp} can also be defined as the Chern-Simons action \eqref{CSaction} with level $k=\frac{1}{4G_N}$ and connection
\begin{equation}\label{Adesp}
A= e^a P_a + \omega^a J_a + \psi^{\alpha}\cG_{\alpha} + \eta^{\alpha} \cR_{\alpha}\,.
\end{equation}
The supertrace in this case reads
\begin{equation}\label{Strdespotic}
\Str(J_aP_b) = \eta_{ab} \,,\qquad \Str(\cG_{\alpha}\cR_{\beta}) = C_{\alpha\beta} \,.
\end{equation}
The action is invariant under supersymmetry off-shell and the supersymmetry transformation laws are
\begin{subequations}
\begin{align}
\delta e^a_\mu & = -\frac12 \left(\bar{\epsilon} \Gamma^a \eta_\mu  + \bar{\zeta} \Gamma^a \psi_\mu \right) \, && \delta \omega^a_\mu   = -\frac12 \bar{\epsilon} \Gamma^a \psi_\mu  \,, \\
\delta \psi_\mu  & = D_\mu \epsilon && \delta \eta_\mu   = D_\mu \zeta +  \frac12 e^a_\mu  \Gamma_a \epsilon\,. \label{despspinor}
\end{align}
\end{subequations}
The commutator of two supersymmetry transformation closes as
\begin{align}
[\delta_{\rm susy}(\epsilon_1,\zeta_1),&\,\delta_{\rm susy}(\epsilon_2,\zeta_2)] =  \; \delta_{\rm g.c.}(\xi^{\nu} = -\tfrac12 (\bar\epsilon_2 \Gamma^{\nu}\zeta_1 + \bar{\zeta}_2 \Gamma^\nu \epsilon_1)) \\ \nonumber 
& + \delta_{\rm  Lor} (\lambda^a = - \xi^{\nu}\omega_{\nu}{}^a - \tfrac12 \bar{\epsilon}_2 \Gamma^a \epsilon_1) + \delta_{\rm susy}(\epsilon = - \xi^{\nu}\psi_{\nu}, \zeta = - \xi^{\nu}\eta_{\nu})\,,
\end{align}
modulo the field equations.

\subsection{Another basis for the algebras}
In the remainder of the paper, we will impose boundary conditions for both flat supergravity theories obtained in this section and discuss the asymptotic and exact Killing spinors. In order to do so, it is convenient to consider a basis where the Chern-Simons connections are naturally written in BMS gauge. 
In both cases we will define
\begin{equation}
L_n = U^a{}_n J_a\,, \qquad M_n = U^a{}_n P_a \,, \qquad \text{with} \;n = -1,0,+1
\end{equation}
for some invertible matrix $U^a{}_n$ defined in \eqref{Umat} as well as redefining all fermionic generators as
\begin{equation}
\cQ_\alpha \to \frac{1}{\sqrt{2}} Q_{-\alpha}
\end{equation}
This map effectively changes the tangent space metric $\eta_{ab}$ to $\gamma_{nm}$ defined as
\begin{equation}\label{gammadef}
\gamma_{nm} = \left( \begin{array}{ccc} 0 & 0 & -2 \\ 0 & 1 & 0 \\ -2 & 0 & 0 \end{array}\right)\,.
\end{equation}
We refer the reader to appendix \ref{sec:conventions} for more details, specifically for our conventions for the gamma matrices in this basis.
Performing this change of basis in the democratic algebra \eqref{democratic2} leads to
\begin{subequations}
\label{democratic}
\begin{align}
[L_n, L_m] & = (n-m)L_{m+n}\,, &&& [L_n, M_m] & = (n-m)M_{m+n}\,, &&& [M_n, M_m] & = 0\,, \\
[L_n,Q^{\pm}_{\alpha}] & = (\tfrac{n}{2}-\alpha)Q^{\pm}_{\alpha+n}\,, &&& [M_n,Q^{\pm}_{\alpha}] & = 0\,, \\
\{Q^{\pm}_{\alpha}, Q^{\pm}_{\beta} \} & = M_{\alpha+\beta}\,, &&& \{Q^{\pm}_{\alpha}, Q^{\mp}_{\beta} \} & = 0\,,
\end{align}
\end{subequations}
while the despotic algebra \eqref{despoticalg} now reads:
\begin{subequations}
\begin{align}\label{despotic}
[L_n, L_m] & = (n-m)L_{m+n}\,, &&& [L_n, M_m] & = (n-m)M_{m+n}\,, \quad [M_n, M_m] = 0\,, \\
[L_n,G_{\alpha}] & = (\tfrac{n}{2}-\alpha)G_{\alpha+n}\,, &&& [L_n,R_{\alpha}] & =  (\tfrac{n}{2}-\alpha)R_{\alpha+n} = [M_n,G_{\alpha}]\,, \\
\{G_{\alpha}, G_{\beta} \} & =  L_{\alpha+\beta}\,, &&& \{G_{\alpha}, R_{\beta} \} & =  M_{\alpha+\beta}\,, \\
 \{R_{\alpha},R_{\beta} \} &  = 0 = [M_n, R_{\alpha}] \,.
\end{align}
\end{subequations}
An interesting observation at this point is that on top of the $\bZ_2$ grading of supersymmetry, it is possible to equip the bosonic sector of both algebras with an additional $\bZ_2$ grading \cite{Gary:2014ppa} under which the $L_n$ generators are even and the $M_n$ odd. The fermionic sector of the despotic case is naturally endowed with the same grading, where $G_\alpha$ ($R_\alpha$) is even (odd), while the fermionic subalgebra \eqref{democratic} is not.

\section{Asymptotic symmetry algebras}\label{sec:bcs}

In this section we will propose boundary conditions and compute the asymptotic symmetry algebra for the two flat space supergravity theories defined in the last section. The asymptotic algebras are highly dependent on the boundary conditions chosen, and hence changing the latter can lead to different algebras. We will focus on specific boundary conditions for the metric at asymptotic lightlike infinity, which can be conveniently written down in a Chern-Simons form following \cite{Barnich:2014cwa}. We find that the asymptotic symmetry algebras are two different $\cN=2$ extensions of the BMS$_3$ algebra, isomorphic to the `homogeneous' and the `inhomogeneous' Supergalilean conformal algebras found in \cite{Bagchi:2016yyf} in the context of tensionless superstrings.   

\subsection{Democratic}
Three dimensional asymptotically flat metrics can be described by a metric in BMS gauge \cite{Barnich:2012aw} with Eddington-Finkelstein coordinates $(u,r,\varphi)$.
\begin{equation}\label{BMSmet}
\extd s^2 =  \gamma_{nm} e^n e^m =  \cM\, \extd u^2 - 2 \extd u \extd r + \cN \, \extd u \extd \varphi + r^2 \extd\varphi^2\,.
\end{equation}
A simple parametrization leading to this metric is to take the Chern-Simons connection \eqref{Adem} in the basis of \eqref{democratic} and in a radial gauge, where the radial dependence $r$ is completely captured by a group element $b$
\begin{equation}\label{democraticbc1}
A = b^{-1}(\extd+a)b\,, \qquad b= \exp\left(\tfrac{r}{2}M_{-1}\right)\,.
\end{equation}
Here $a = a(u,\vp)$ only has legs in the $u,\vp$ components, i.e. $a = a_u du+a_\vp d\vp$, with:
\begin{align}\label{democraticbc2}
a_u & = M_{+1} - \frac{\cM}{4} M_{-1}\,. \\
a_{\vp} & = L_{+1} - \frac{\cM}{4}L_{-1} - \frac{\cN}{4} M_{-1} + \tfrac14\, \sum_{i=+,-}\psi^i Q^{i}_{-}\,.
\label{democraticbc3}
\end{align}
Here $\cM, \cN$ and the Grassmann-valued spinor components $\psi^{i}$ are functions of $(u,\vp)$ only and the index $i$ labels the two fermionic charges. The field equations constrain these functions as
\begin{align}
\partial_u \cM & = 0\,, & \partial_u \cN & =  \cM' \,, &
\partial_u \psi^{i} & = 0\,.
\end{align}
Here primes denote derivatives with respect to $\vp$. These equations are solved by
\begin{equation}
\cM  = \cM(\vp)\,, \quad \cN = \cJ(\vp) + u \cM'(\vp)\,, \quad \psi^{i} = \Psi^{i}(\vp)\,.
\end{equation}
We will now look for the most general gauge transformation which preserves the form \eqref{democraticbc1}-\eqref{democraticbc3}, i.e. we are looking for the gauge parameter $\lambda$ which satisfy $\delta a = \extd\lambda + [a,\lambda]$. If we parametrize $\lambda$ as
\begin{equation}
\lambda = \xi^n M_n + \chi^n L_n + \epsilon_{i}^{\alpha} Q^{i}_{\alpha}
\end{equation} 
Then from solving $\delta a_{\vp} = \partial_{\vp}\lambda + [a_{\vp},\lambda]$ we find the following 6 relations among the gauge parameters appearing in $\lambda$:
\begin{subequations}
\begin{align}
\chi^0 & = - \chi^+{}'\,,  & \chi^- & = \frac12 \chi^+{}'' - \frac{\cM}{4}\chi^+\,,   \\
\xi^0 & = - \xi^+{}'\,, & \xi^- & = \frac12 \xi^+{}'' -\frac{\cM}{4}\xi^+ - \frac{\cN}{4} \chi^+ - \frac18 \psi^{i} \epsilon_{i}^{+} \,, \\
\epsilon^{-}_{i} & = - \partial_{\vp}\epsilon^{+}_{i} + \frac14\,\psi^{i} \chi^+\,.
\end{align}
\end{subequations}
Hence, the variations of the functions appearing in \eqref{democraticbc3} can be expressed in terms of the four, up to now arbitrary, gauge parameters $\chi^+, \xi^+, \epsilon^{+}_{i}$.
From the transformation of the $u$-component of the connection, i.e. $\delta a_{u} = \partial_{u}\lambda + [a_{u},\lambda]$, we find that 
\begin{align}
\partial_u \chi^+ & = 0\,, & \partial_u \epsilon^{+}_{i} & = 0\,, & \partial_u \xi^+ & = \chi^+{}' \,.
\end{align}
This fixes the $u$ behavior of the gauge parameters. We take them to be:
\begin{equation}
\chi^+ = Y(\vp) \,, \qquad \xi^+ = T(\vp) + u\, Y'(\vp)\,, \qquad \epsilon^{+}_{i} = \epsilon_{i}(\vp) 
\end{equation}
The variation of the functions $\cM, \cJ$ and $\Psi^{i}$ under the asymptotic symmetry transformations is then given by:
\begin{subequations}
\label{vardemocratic}
\begin{align}
\delta \cM & = - 2 Y''' + 2 \cM Y' + \cM' Y\,, \\
\delta \cJ & = - 2 T''' + 2 \cM T' + \cM' T + 2 \cJ' Y + \cJ' Y + \frac32 \epsilon_{\pm}'\Psi^{\pm} + \frac12 \epsilon_{\pm} \Psi^{\pm}{}' \,, \\
\delta \Psi^{i} & = - 4\,\epsilon_{i}'' + \frac32 Y' \Psi^{i} + Y \Psi^{i}{}' + \cM \epsilon_{i}\,. \label{demvarspin}
\end{align}
\end{subequations}
We can now compute the variation of the asymptotic charges directly from the background independent result \cite{Banados:1994tn}
\begin{equation}
\label{delQ}
\delta Q[\lambda] = - \frac{k}{2\pi} \int \Str\left(\lambda \, \delta a_{\vp} \right) \extd\vp\,.
\end{equation}
Upon using the supertrace \eqref{Strdemocratic} we find that the charges are trivially integrated (assuming that the variation of $T, Y$ and $\epsilon_{i}$ vanishes) to give
\begin{equation}
Q[T, Y,\epsilon_{\pm}] = -\frac{k}{4\pi} \int \left(T \cM + Y \cJ + \epsilon_{i} \Psi^{i} \right) \extd\vp
\end{equation}
The Poisson bracket algebra of these charges can then readily be computed using $\{Q[\lambda_1],Q[\lambda_2]\}_{P.B.} = \delta_{\lambda_1}Q[\lambda_2]$. We represent the result in terms of the Fourier components:
\begin{subequations}
\begin{align}
\cJ_n & = \frac{k}{4\pi} \int \extd\vp \, e^{in\vp} \cJ(\vp)\,, \\
\cM_n & = \frac{k}{4\pi} \int \extd\vp \, e^{in\vp} \cM(\vp)\,, \label{Mdef} \\
\Psi^{\pm}_r & = -\frac{k}{4\pi} \int \extd\vp \, e^{ir\vp} \Psi^{\pm}(\vp)\,. 
\end{align}
\end{subequations}
After converting the Poisson brackets to commutators via $i\{,\}_{P.B.} \to [,]$ and to anti-commutators $\{,\}_{P.B.} \to \{,\}$ for the fermionic charges, the asymptotic symmetry algebra reads:
\begin{subequations}
\label{democraticASA}
\begin{align}
[\cJ_n, \cJ_m] & = (n-m)\cJ_{m+n} + \frac{c_L}{12} n^3 \delta_{m+n,0}\,, \\
[\cJ_n, \cM_m] & = (n-m)\cM_{m+n} + \frac{c_M}{12}n^3 \delta_{m+n,0}\,, \\
[\cJ_n,\Psi^{\pm}_{r}] & = (\tfrac{n}{2}-r)\Psi^{\pm}_{r+n}\,, \\
\{\Psi^{\pm}_{r}, \Psi^{\pm}_{s} \} & = \cM_{r+s} + \frac{c_M}{6} r^2  \delta_{r+s,0}\,, \label{PsiPsiM}\\
\{\Psi^{\pm}_{r},\Psi^{\mp}_{s} \} &  = 0 = [\cM_n, \cM_m] = [\cM_n, \Psi^{\pm}_{r}] \,.
\end{align}
\end{subequations}
where $c_L=0$ and $c_M = 12k = \frac{3}{G_N}$. From this algebra the standard form of the central terms $\frac{c_L}{12}(n^3-n)$, $\frac{c_M}{12}(n^3-n)$ and $\frac{c_M}{6}(r^2- \frac14)$ can be obtained by shifting the zero modes $\cJ_0 \to \cJ_0 + \frac{c_L}{24}$ and  $\cM_0 \to \cM_0 + \frac{c_M}{24}$.

This is the $\cN =2$ super BMS$_3$ algebra found in \cite{Casali:2016atr, Banerjee:2016nio}, called the `homogeneous' super Galilean conformal algebra in \cite{Bagchi:2016yyf}, where this algebra was obtained from an ultra-relativistic contraction of the two dimensional $\cN = (1,1)$ superconformal algebra.

\subsection{Despotic}
We will now focus on the other (despotic) flat space limit of the last section. Once again, we take the connection to be in a radial gauge
\begin{equation}\label{despoticbc1}
A = b^{-1}(d+a)b\,, \qquad b= \exp\left(\tfrac{r}{2}M_{-1}\right)\,,
\end{equation}
where $a = a(u,\vp)$ only has legs in the $u,\vp$ components but are now given by:
\begin{align}\label{despoticbc2}
a_u & = M_{+1} - \frac{\cM}{4} M_{-1} +\tfrac14 \psi R_{-}\,. \\
a_{\vp} & = L_{+1} - \frac{\cM}{4}L_{-1} - \frac{\cN}{4} M_{-1} + \tfrac14\,\psi G_{-} + \tfrac14\,\eta R_{-}\,.
\label{despoticbc3}
\end{align}
The fields $\cM, \cN$ and the Grassmann-valued spinor components $\psi$ and $\eta$ are functions of $(u,\vp)$ only and the field equations for this solution now constrain the functions as
\begin{align}
\partial_u \cM & = 0\,, & \partial_u \cN & =  \cM' \,,\\
\partial_u \psi & = 0\,, & \partial_u \eta & =   \psi'\,.
\end{align}
The solution is parametrized by the functions $\cM(\vp),\cJ(\vp), \Psi(\vp)$ and $\Theta(\vp)$ defined as:
\begin{equation}
\cM  = \cM(\vp)\,, \quad \cN = \cJ(\vp) + u\, \cM'(\vp)\,, \quad \psi = \Psi(\vp)\,, \quad \eta = \Theta(\vp) + u\, \Psi'(\vp)\,.
\end{equation}
A gauge transformation $\delta a = \extd\lambda + [a,\lambda]$ which preserves the form \eqref{despoticbc1}-\eqref{despoticbc3} can be parametrized by a parameter $\lambda$ of the form:
\begin{equation}
\lambda = \xi^n M_n + \chi^n L_n + \epsilon^{\alpha} G_{\alpha} + \zeta^{\alpha} R_{\alpha}\,.
\end{equation} 
From preserving the form of $\delta a_{\vp}$ we find that 6 of the gauge parameters in $\lambda$ are constrained to be:
\begin{align}
\chi^0 & = - \chi^+{}'\,, &
\chi^- & = \frac12 \chi^+{}'' - \frac{\cM}{4}\chi^+ - \frac18 \psi \epsilon^+\,,  \\
\xi^0 & = - \xi^+{}'\,, &
\xi^- & = \frac12 \xi^+{}'' -\frac{\cM}{4}\xi^+ - \frac{\cN}{4} \chi^+ - \frac18 \psi \zeta^+ - \frac18 \eta \epsilon^+ \,, \\
\epsilon^- & = - \epsilon^+{}' + \frac14\,\psi \chi^+\,, & 
\zeta^- & = - \zeta^+{}' + \frac14\,\eta \chi^+ + \frac14\,\psi \xi^+\,. 
\end{align}
The four, up to now arbitrary, gauge parameters are $\chi^+, \xi^+, \epsilon^+$ and $\zeta^+$.
From the preservation of $\delta a_{u}$ we find that their $u$ dependence should take the form
\begin{align}
\partial_u \chi^+ & = 0\,, & \partial_u \epsilon^{+} & = 0\,, \\
\partial_u \xi^+ & = \chi^+{}'\,, & \partial_u \zeta^+ & = \epsilon^+{}' \,.
\end{align}
We solve for the $u$-behavior explicitly by writing:
\begin{equation}
\chi^+ = Y(\vp) \,, \qquad \xi^+ = T(\vp) + u\, Y'(\vp)\,, \qquad \epsilon^+ = \epsilon(\vp) \,, \qquad \zeta^+ = \zeta(\vp) + u\, \epsilon'(\vp)\,.
\end{equation}
The variation of the functions $\cM, \cJ, \Psi$ and $\Theta$ under the asymptotic symmetry transformations is then given by:
\begin{subequations}
\label{vardespotic}
\begin{align}
\delta \cM & = - 2 Y''' + 2 \cM Y' + \cM' Y + \frac32 \epsilon' \Psi +\frac12 \epsilon \Psi'\,, \\
\delta \cJ & = - 2 T''' + 2 \cM T' + \cM' T + 2 \cJ' Y + \cJ' Y + \frac32 \zeta'\Psi + \frac12 \zeta \Psi' + \frac32 \epsilon'\Theta + \frac12 \epsilon \Theta' \,, \\
\delta \Psi & = - 4\,\epsilon'' + \frac32 Y' \Psi + Y \Psi' + \cM \epsilon\,, \\
\delta \Theta & = - 4\,\zeta''  + \frac32 Y' \Theta + Y \Theta' + \frac32 T' \Psi + T \Psi' + \cM\zeta +\cJ\epsilon\,.
\end{align}
\end{subequations}
The expression for the variation of the charges \eqref{delQ} (now evaluated with the supertrace \eqref{Strdespotic}) is once again trivially integrated and gives the finite and conserved result:
\begin{equation}
Q[T, Y,\epsilon, \zeta] = -\frac{k}{4\pi} \int \left(T \cM + Y \cJ + \epsilon \Theta + \zeta \Psi \right) \extd\vp \,.
\end{equation}
In terms of the Fourier components
\begin{subequations}
\begin{align}
\cJ_n & = \frac{k}{4\pi} \int \extd\vp \, e^{in\vp} \cJ(\vp)\,, \\
\cM_n & = \frac{k}{4\pi} \int \extd\vp \, e^{in\vp} \cM(\vp)\,, \\
\Theta_r & = -\frac{k}{4\pi} \int \extd\vp \, e^{ir\vp} \Theta(\vp)\,, \\
\Psi_r & = -\frac{k}{4\pi} \int \extd\vp \, e^{ir\vp} \Psi(\vp)\,, 
\end{align}
\end{subequations}
the asymptotic symmetry algebra is in this case:
\begin{subequations}
\label{despoticASA}
\begin{align}
[\cJ_n, \cJ_m] & = (n-m)\cJ_{m+n} + \frac{c_L}{12} n^3 \delta_{m+n,0}\,, \\
[\cJ_n, \cM_m] & = (n-m)\cM_{m+n} + \frac{c_M}{12} n^3 \delta_{m+n,0}\,, \\
[\cM_n, \cM_m] & = 0\,, \\
[\cJ_n,\Theta_{r}] & = (\tfrac{n}{2}-r)\Theta_{r+n}\,, \\
[\cJ_n,\Psi_{r}] & =  (\tfrac{n}{2}-r)\Psi_{r+n} = [\cM_n,\Theta_{r}]\,, \\
\{\Theta_{r}, \Theta_{s} \} & =  \cJ_{r+s} + \frac{c_L}{6}r^2 \delta_{r+s,0}\,, \\
\{\Theta_{r}, \Psi_{s} \} & = \cM_{r+s} + \frac{c_M}{6}r^2 \delta_{r+s,0}\,, \\
\{\Psi_{r},\Psi_{s} \} &  = 0 = [\cM_n, \Psi_{r}] \,.
\end{align}
\end{subequations}
This symmetry algebra is isomorphic to the `inhomogeneous' $\cN=2$ Supergalilean conformal algebra of \cite{Mandal:2010gx,Bagchi:2016yyf} with $c_L = 0$ and $c_M = 3/G_N$. It is possible to obtain this algebra from both a non-relativistic \cite{Mandal:2010gx} and an ultra-relativistic limit \cite{Bagchi:2016yyf} of the $\cN=(1,1)$ superconformal algebra. However, both limits have peculiar properties and hence we will briefly elaborate on them now. The $\cN =(1,1)$ superconformal algebra is
\begin{subequations}
\label{superconf}
\begin{align}
[\cL^{\pm}_n, \cL^{\pm}_m] & = (n-m)\cJ_{m+n} + \frac{c^{\pm}}{12}(n^3-n)\delta_{n+m,0}\,, \\
[\cL^{\pm}_n, \cQ^{\pm}_{r}] & =  (\tfrac{n}{2}-r)\cQ^{\pm}_{n+r}  \\
\{\cQ^{\pm}_{r}, \cQ^{\pm}_{s} \} & = 2\cL^{\pm}_{r+s} + \frac{c^{\pm}}{6}\left(r^2+\frac14\right) \delta_{r+s,0}\,.
\end{align}
\end{subequations}
The non-relativistic limit is defined by the $\ell \to \infty$ limit of the above algebra with
\begin{align}
\cJ_n & = \cL^+_n+\cL^-_n & \cM_n & = \frac{1}{\ell}(\cL^+_n-\cL^-_n) \\
 \Theta_{r}  & =  \sqrt{2}(\cQ^+_{r} + \cQ^-_{r}) \,, & \Psi_{r}& = \frac{\sqrt{2}}{\ell} (\cQ^+_{r} -  \cQ^-_{r})\,.
\end{align}
This gives the algebra \eqref{despoticASA} with $c_L = c^+ + c^-$, $c_M =  \frac{1}{\ell}(c^+-c^-)$ and with shifted zero-modes. Because of the vanishing of $c_L$, this limit is not entirely satisfactory as, in this case, it enforces one of the chiral sectors to have negative central charge.

The other limit, called the ultra-relativistic limit, is obtained by taking $\ell \to \infty$, while the generators scale as
\begin{align}
\cJ_n & = \cL^+_n - \cL^-_{-n} & \cM_n & = \frac{1}{\ell} (\cL^+_n + \cL^-_{-n}) \\
\Theta_{r} & =  \sqrt{2}(\cQ^+_{r} - i \cQ^-_{-r}) \,, &  \Psi_{r} & = \frac{\sqrt{2}}{\ell} (\cQ^+_{r} +  i \cQ^-_{-r})\,. \label{URlimit}
\end{align}
Now the central charges scale as $c_L = c^+ - c^-$ and $c_M =  \frac{1}{\ell}(c^+ + c^-)$, so we do not encounter the problem of having to start with negative central charges. However, this scaling limit is still not well-defined as the Hermitian conjugates of the fermionic generators are ill-defined in the limit. If we assume that the superconformal generators $\cQ^{\pm}$ are Hermitian $(\cQ^\pm_r)^\dagger = \cQ^\pm_{-r}$, then due to the factor of $i$ in the definitions \eqref{URlimit} we have
\begin{equation}\label{adjointscale}
\Psi^\dagger_{r}=\ell\, \Theta_{-r}\,,\qquad \Theta^\dagger_{\alpha}= \frac{1}{\ell}\,\Psi_{-r}\,.
\end{equation} 
Hence the adjoint either diverges or vanishes in the $\ell \to \infty$ limit. A possible resolution in this case would be to start with a $\cN=(1,1)$ superconformal algebra, but take $\cQ^-$ to be anti-Hermitian $(\cQ^-_r)^\dagger = -\cQ^-_{-r}$. The ultra-relativistic limit in this case would give
\begin{equation}
\label{eq:adjoint_good}
(\Psi_r)^\dagger= \Psi_{-r}\;,\qquad (\Theta_r)^\dagger= \Theta_{-r}\;.
\end{equation}
Unlike \eqref{adjointscale}, these relations are consistent with the algebra \eqref{despoticASA} after taking $\ell \to\infty$ and hence these should be the correct Hermitian conjugates, although it is not clear how this relation would follow from the point of view of a limit of unitary representations of the $\cN=(1,1)$ superconformal algebra. This problem is the fermionic counterpart of the known subtleties in defining unitary highest-weight representations of the BMS$_3$ algebra as a limit from those of the 2D conformal algebra \cite{Bagchi:2009pe,Grumiller:2014lna}. In fact, by analogy with the Poincare algebra, one  expects the unitary representations of the BMS$_3$ algebra to be induced representations. These indeed follow from the ultra-relativistic limit of the highest-weight representations of the Virasoro algebra \cite{Campoleoni:2015qrh,Campoleoni:2016vsh}. It would be interesting to investigate this relation for the despotic super-BMS$_3$ algebra, but for the moment we will not worry about this and simply define the Hermitian conjugates as \eqref{eq:adjoint_good}.


\section{Killing spinors and energy bounds}\label{sec:Killspin}
In this section we will discuss some properties of the two flat space supergravity theories. We will first look at the energy bounds imposed by the algebra upon assuming Neveu-Schwarz boundary conditions for the fermions. Then we solve the asymptotic Killing spinor equation and the exact Killing spinor equation for purely bosonic zero mode solutions in both of the flat supergravity theories. 

\subsection{Energy bounds}
It is possible to derive an energy bound from the algebras \eqref{democraticASA} and \eqref{despoticASA} along the lines of \cite{Deser:1977hu,Abbott:1981ff,Coussaert:1994jp}. The wedge subalgebra for both limits is spanned by the set $\cM_n, \cJ_n$ with $n = -1,0,+1$ while the index for fermionic generators take values $r= \pm 1/2$ and $r=0$ for the Neveu-Schwarz and Ramond fermionic boundary conditions respectively.

\subsubsection{Democratic}
The democratic case has as wedge subalgebra the usual $\cN=2$ super-Poincar\'e algebra. In the quantum theory, their bracket \eqref{PsiPsiM} becomes an anticommutator to lowest order in $\hbar$ and the quantum generator $\cM_0$ associated to time translations is bounded as
\begin{equation}\label{dembound}
\cM_0 = \frac{1}{2}\sum_{i=+,-}\left(\Psi^i_{1/2}\Psi^i_{-1/2} + \Psi^i_{-1/2}\Psi^i_{1/2}\right) - \frac{k}{2} \geq - \frac{1}{8G}\,.
\end{equation} 
This bound is saturated by the Minkowski vacuum, which has all charges vanishing, except $\cM = -1$ (and hence, by  \eqref{Mdef}, $\cM_0 = -1/8G$).

When imposing Ramond boundary conditions on the fermions, the modes $\Psi_{r}^{\pm}$ have integer $r$ and the bound on the quantum generators simply becomes $\cM_0 = \sum_i(\Psi^i_{0})^2 \geq 0$. This bound is saturated by the null orbifold solution of \cite{Horowitz:1990ap}, which has all modes vanishing.

\subsubsection{Despotic}

In the despotic algebra \eqref{despoticASA}, very unusually, the time translation generator $\cM_0$ appears on the r.h.s of the mixed anti-commutator of the fermions and hence, in the quantum theory, we have:
\begin{align}
\label{desbound}
\cM_0 &=\tfrac12\,\{\Theta_{1/2},\Psi_{-1/2}\}+\tfrac12\,\{\Theta_{-1/2},\Psi_{1/2}\} - \frac{1}{8G}
\nonumber\\
&= \tfrac12\,(\Theta_{1/2}\Psi_{-1/2} + \Psi_{-1/2}\Theta_{1/2}+ \Theta_{-1/2}\Psi_{1/2} + \Psi_{1/2}\Theta_{-1/2}) - \frac{1}{8G}\,,
\end{align} 
Since the adjoints in this case are given by \eqref{eq:adjoint_good} it is not immediate to derive an energy bound for supersymmetric quantum states, but we will need some auxiliary results which are peculiar to this case. Specifically, a new bound on the expectation value of the quantum operator $\cJ_0$ exists:
\begin{equation}
\cJ_0=\Theta_{1/2}\,\Theta_{-1/2}+\Theta_{-1/2}\,\Theta_{1/2}\ge 0\;.
\end{equation}
So the quantum generator $\cJ_0$ is bounded from below and its expectation value must be positive. We can exploit this atypical result to obtain a bound on the energy as follows. Call $\cQ_r=\tfrac1{\sqrt{2}}(\Theta_{r}+\Psi_{r})$ and $\cQ_r^\dagger=\tfrac1{\sqrt{2}}(\Theta_{-r}+\Psi_{-r})$. Then we will have in the quantum theory
\begin{equation}
0 \le \cQ_{1/2}\,\cQ_{1/2}^\dagger+\cQ_{1/2}^\dagger\,\cQ_{1/2} = \frac12\,\cJ_0+\cM_0+\frac{1}{8G}
\end{equation}
where we used the fact that $\Psi$ anticommutes with itself. From the above equality, the bound on the energy $\cM_0$ reads:
\begin{equation}
\cM_0\ge -\frac1{8G}\,(4G\,\cJ_0+1)
\end{equation}
There is another bound on the energy which can be obtained by considering the combination $\tilde{\cQ}_r=\tfrac1{\sqrt{2}}(\Theta_{r}-\Psi_{r})$ and $\tilde{\cQ}_r^\dagger=\tfrac1{\sqrt{2}}(\Theta_{-r}-\Psi_{-r})$, i.e.:
\begin{equation}
0 \le \tilde{\cQ}_{1/2}\,\tilde{\cQ}_{1/2}^\dagger+\tilde{\cQ}_{1/2}^\dagger\,\tilde{\cQ}_{1/2} = \frac12\,\cJ_0-\cM_0-\frac{1}{8G}\,.
\end{equation}
So, the despotic supersymmetry algebra forces the energy of states to be bound both from below and above:
\begin{equation}\label{despbound}
-\frac12\cJ_0\le\cM_0 +\frac1{8G}\le \frac12\cJ_0\,.
\end{equation}
Surprisingly, $|\cM_0|$ is bound by $\cJ_0$, instead of the other way around, which would lead to a customary BPS bound. Hence it seems as if the roles of $\cJ_0$ and $\cM_0$ are interchanged in the despotic theory. 

The above bound is still saturated by the Minkowski vacuum, for which $\cJ$ and hence $\cJ_0$ vanishes. However from this argument alone it is insufficient to conclude that the Minkowski vacuum has the lowest energy. In principle \eqref{despbound} could allow for a state with $\cJ_0\neq 0$ and with lower energy than Minkowski space. Below we will show that the solution of the exact Killing spinor equation is only globally well-defined if $\cJ=0$ and hence if such a state does exist, then it will not be supersymmetric and hence it can not saturate the bound \eqref{despbound}.\\
For Ramond boundary conditions on the fermions we can use similar arguments as above to write
\begin{align}\label{despRamond}
\cM_0 &= \Theta_0 \Psi_0 + \Psi_0 \Theta_0 = 2 (\cQ_0)^2 - (\Theta_0)^2 - (\Psi_0)^2 = 2 (\cQ_0)^2 - \cJ_0 \geq - \cJ_0
\nonumber\\
\cM_0 &= \Theta_0 \Psi_0 + \Psi_0 \Theta_0 = - 2 (\tilde{\cQ}_0)^2 + (\Theta_0)^2 + (\Psi_0)^2 =- 2 (\tilde{\cQ}_0)^2 + \cJ_0 \leq  \cJ_0\,,
\end{align}
So again the energy is bound to be:
\begin{equation}
-\cJ_0\le \cM_0 \le \cJ_0
\end{equation}
The null orbifold with $\cM =0$ and $\cJ=0$ saturates this bound, but once again, for states with non-zero $\cJ_0$ the bound on the energy is weaker.  

\subsection{Asymptotic Killing spinors}
We will now switch our attention to the asymptotic supersymmetry transformations which leave the purely bosonic solution invariant, i.e. we will solve the asymptotic Killing spinor equations for both flat limits.

\subsubsection{Democratic}
The asymptotic Killing spinor equation is obtained by setting the fermionic charges to zero and demanding that the asymptotic symmetry transformation \eqref{demvarspin} doesn't generate any new fermionic charge. In other words, we should solve: 
\begin{equation}
\delta \Psi^{i} = - 4 \epsilon_{i}'' + \cM \epsilon_{i} = 0
\end{equation}
The solution for $\cM \neq 0$ is:
\begin{equation}
\epsilon_{i} = c_1^{i} e^{-\frac{\sqrt{\cM}}{2}\varphi} + c_2^{i} e^{\frac{\sqrt{\cM}}{2}\varphi}\,.
\end{equation}
This is globally well-defined when $\cM = - n^2$, where we take $n>0$ without loss of generality. The bound \eqref{dembound} is saturated for $n=1$, which corresponds to the Minkowski ground state and violated for $n>1$, which are angular defect solutions.  

When $\cM = 0$, the solution is
\begin{equation}
\epsilon_{i} = E_0^{i} + \varphi F_0^{i}\,.
\end{equation}
This is single valued only when $F_0^{i}=0$ and hence the null orbifold of \cite{Cornalba:2002nv,Cornalba:2002fi} with $\cM=0=\cJ$ has a single solution per asymptotic Killing spinor equation. The situation in this case is a straight forward extension of the $\cN=1$ flat supergravity theory described in \cite{Barnich:2014cwa}.

\subsubsection{Despotic}
In the despotic case, the asymptotic Killing spinor equations become coupled and they read explicitly:
\begin{equation}
\delta \Psi = - 4 \epsilon'' + \cM \epsilon = 0 \qquad \delta \Theta = - 4 \zeta'' + \cM\zeta + \cJ \epsilon = 0\,.
\end{equation}
By inspection it is easy to see that the solution for $\cM \neq 0$, $\cJ = 0$ is the same as in the last section. When $\cJ\neq0$, we obtain:
\begin{equation}
\epsilon = c_1 e^{-\frac{\sqrt{\cM}}{2}\varphi} + c_2 e^{\frac{\sqrt{\cM}}{2}\varphi}\,. \qquad
\zeta = \left(c_3 - \frac{c_1 \cJ\, \varphi }{4\sqrt{\cM}} \right)  e^{-\frac{\sqrt{\cM}}{2}\varphi} +
\left(c_4 + \frac{ c_2 \cJ\,  \varphi}{4 \sqrt{\cM}}  \right) e^{\frac{\sqrt{\cM}}{2}\varphi}
\end{equation}
Due to the periodicity in $\varphi$ this is only globally well-defined when $\cJ = 0$ (and $\cM = - n^2$, as before). This once again singles out the Minkowski vacuum with $\cM = -1$ and $\cJ=0$. Note that in contrast to the previous case, here $\cJ=0$ is required explicitly by demanding the Killing spinor be globally well-defined and hence the energy bound \eqref{despbound} becomes a hard equality $\cM_0 = -\frac{1}{8G}$, exactly the value of the Minkowski vacuum.

When $\cM = 0$, the solution is simply
\begin{equation}
\epsilon = E_0 + \varphi E_1\,, \qquad \zeta = Z_0 + \varphi Z_1 + \frac{\cJ E_0}{8} \varphi^2 + \frac{\cJ E_1}{24} \varphi^3\,.
\end{equation}
This is single valued only when $E_1=0=Z_1$ and $\cJ =0$. Once again, this gives a single periodic solution per asymptotic Killing spinor for the null orbifold solution.

\subsection{Exact Killing spinors}
We will now focus on the exact Killing spinors of the purely bosonic zero-mode solutions. These solutions are described by the metric:
\begin{equation}
\extd s^2 = \cM\, \extd u^2 - 2 \extd u \extd r + \cJ \, \extd u \extd \varphi + r^2 \extd\varphi^2\,,
\end{equation}
with constant $\cM$ and $\cJ$. For both flat space limits, we will choose a frame where the vielbein is parameterized by
\begin{equation}
e = \frac12 M_{-1}\, \extd r -  \left(\frac{\cJ}{4}\, M_{-1} - r\, M_{0}\right) \,\extd \varphi + \left(M_{+1} - \frac{\cM}{4}M_{-1}\right) \, \extd u\,,
\end{equation}
and the spin connection is
\begin{equation}
\omega = \left(L_{+1} - \frac{\cM}{4}L_{-1} \right)\, \extd \varphi 
\end{equation}
In the following sections we are using the $\Gamma$-matrices in a basis suited for the bosonic $isl(2)$ algebra which are denoted by tilde's (see appendix \ref{sec:conventions}).

\subsubsection{Democratic}

The exact Killing spinor equations are:
\begin{equation}
D\epsilon =0 \,, \qquad D\zeta =0\,.
\end{equation}
Here $D  = d + \omega$ with $\omega = \frac12 \omega^n \tilde{\Gamma}_n$.

The equations can be solved by writing $\epsilon = \Lambda^{-1}\epsilon_0$ and $\zeta = \Lambda^{-1}\zeta_0$ for constant spinors $\epsilon_0$ and $\zeta_0$ and
\begin{equation}
\omega = \Lambda^{-1} d \Lambda\,, \qquad \Lambda = \exp\left(\frac12 \left(\tilde{\Gamma}_{+1} - \frac{\cM}{4}\tilde{\Gamma}_{-1} \right)\varphi \right)\,.
\end{equation}
Explicitly, the solution to the Killing spinor equations reads:
\begin{equation}
\epsilon = \left(\begin{array}{cc}
\cosh\left(\frac{\sqrt{\cM}}{2}\varphi \right) & -\frac{\sqrt{\cM}}{2}\sinh\left(\frac{\sqrt{\cM}}{2}\varphi \right) \\
-\frac{2}{\sqrt{\cM}}\sinh\left(\frac{\sqrt{\cM}}{2}\varphi \right) & \cosh\left(\frac{\sqrt{\cM}}{2}\varphi \right)
\end{array}\right)\epsilon_0\,,
\end{equation}
and likewise for $\zeta$. 

As is the case for the asymptotic Killing spinors, the solution is globally defined when $\cM = - n^2$. When $n>0$ there are two independent solutions, while for $n=0$ the $\tilde{\Gamma}_{+1}$ contribution to $\epsilon$ diverges, which can simply be remedied by taking $\epsilon_0 = (0,\epsilon_0^-)$. Hence in this case there is only one independent solution to the exact Killing equation, exactly as for the asymptotic Killing spinor.

\subsubsection{Despotic}
In the case of the despotic flat space limit, finding the exact Killing spinors is a little more involved due to the inhomogeneous term involving $\epsilon$ in the supersymmetry transformation of $\eta$ in \eqref{despspinor}. The equations to solve are now:
\begin{equation}\label{despKillSpin}
D\epsilon = 0 \,, \qquad D\zeta = - \frac12 e^n \tilde{\Gamma}_{n} \epsilon\,.
\end{equation}
The solution for $\epsilon$ remains unchanged and we can construct the solution for $\zeta$ by adding the solution of the homogeneous equation with coordinate dependent coefficients to the solution of the last section. Explicitly, take
\begin{equation}
\zeta = \Lambda^{-1}\zeta_0 + \Lambda^{-1}\chi(x)\,,
\end{equation}
where $\chi(x)$ is a spinor depending on all coordinates $x= \{u,r,\varphi\}$ which we will now determine. The second Killing spinor equation in \eqref{despKillSpin} becomes:
\begin{align}
\Lambda^{-1}\, d\chi & = - \frac12 e^{n}\tilde{\Gamma}_n \epsilon  \nonumber \\
& = - \frac14 \tilde{\Gamma}_{-1}\epsilon\, dr + \frac12\left(\frac{\cJ}{4}\tilde{\Gamma}_{-1} - r\, \tilde{\Gamma}_{0}\right) \epsilon \,d\varphi - \frac12\left(\tilde{\Gamma}_{+1} - \frac{\cM}{4}\tilde{\Gamma}_{-1}\right) \epsilon \,du\,.
\end{align}
This implies that $\chi(x)$ is a solution to these three partial differential spinor equations
\begin{subequations}
\label{chieqn}
\begin{align}
\Lambda^{-1} \partial_r \chi & =  - \frac14 \tilde{\Gamma}_{-1} \epsilon\,, \\
\Lambda^{-1} \partial_{\varphi} \chi & = \frac12\left(\frac{\cJ}{4}\tilde{\Gamma}_{-1} - r \,\tilde{\Gamma}_{0}\right) \epsilon\,,  \\
\Lambda^{-1} \partial_u \chi & = - \frac12\left(\tilde{\Gamma}_{+1}- \frac{\cM}{4}\tilde{\Gamma}_{-1}\right) \epsilon\,.
\end{align}
\end{subequations}
The first equation immediately gives:
\begin{equation}
\Lambda^{-1} \chi = - \frac{r}{4} \tilde{\Gamma}_{-1} \epsilon + \Lambda^{-1}\chi_1(u,\varphi)\,,
\end{equation}
with $\chi_1$ an arbitrary spinor depending only on $u$ and $\varphi$. After using this in the second equation in \eqref{chieqn}, along with the fact that $[\Lambda^{-1}\partial_{\varphi}\Lambda, \tilde{\Gamma}_{-1}] = 2 \tilde{\Gamma}_0$, one can observe that the linear terms in $r$ cancel and we are left with the equation
\begin{equation}
\partial_{\varphi} \chi_1 = \frac{\cJ}{8}\Lambda \tilde{\Gamma}_{-1}\Lambda^{-1}\epsilon_0\,.
\end{equation}
Explicit integration gives the solution
\begin{equation}
\Lambda^{-1}\chi_1(u,\varphi) = \frac{\cJ}{4\cM}\left[ \varphi \left(\frac{\cM}{4}\tilde{\Gamma}_{-1} -  \tilde{\Gamma}_{+1} \right) + \tilde{\Gamma}_0 \right]\epsilon + \Lambda^{-1}\chi_2(u)\,.
\end{equation}
Finally, the $u$ dependent piece is found from the last of equations \eqref{chieqn}
\begin{equation}
\Lambda^{-1} \chi_2(u) = \frac{u}{2} \left(\frac{\cM}{4}\tilde{\Gamma}_{-1} - \tilde{\Gamma}_{+1} \right)\epsilon =  u \partial_{\varphi}\epsilon\,.
\end{equation}
Hence the general solution to the Killing spinor equations \eqref{despKillSpin} is
\begin{align}
\epsilon & = \Lambda^{-1}\epsilon_0\,, \qquad \Lambda = \exp\left(\frac12 \left(\tilde{\Gamma}_{+1} - \frac{\cM}{4}\tilde{\Gamma}_{-1} \right)\varphi \right)\,,\\
\zeta & = \Lambda^{-1}\zeta_0 + \left[ \frac{\cJ}{4\cM}\tilde{\Gamma}_0 + \left(\frac{\cJ}{2\cM} \varphi + u \right)\partial_{\varphi} - \frac{r}{4} \tilde{\Gamma}_{-1}\right] \epsilon\,.
\end{align}
This solution contains terms linear in the coordinates. For single-valuedness of the solution, the term linear in $\varphi$ should vanish, which it does for $\cJ=0$. Hence the presence of an exact Killing spinor in this case explicitly requires $\cJ=0$, while the periodicity requirements remain unchanged with respect to the democratic case (i.e. we still have that $\cM = -n^2$ gives two independent solutions when $n>0$ and one when $n=0$). This shows that there are no supersymmetric solutions with constant $\cJ \neq 0$.

\subsection*{Acknowledgments}
The authors have the pleasure to thank Arjun Bagchi, Daniel Butter, Daniel Grumiller, Sunil Mukhi, Valentin Reys, Jan Rosseel and Bernard de Wit for many useful discussions and illuminating comments.
WM was supported by the FWF project P 27182-N27. IL work was partially supported by a DST Ramanujan Grant.

\appendix
\section{Conventions}
\label{sec:conventions}
Throughout the paper we follow conventions where the Levi-Civita symbol is completely anti-symmetric with $\epsilon_{012} = +1$ and the tangent space metric is the 3D Minkowski metric
\begin{equation}
\eta_{ab} = \left( \begin{array}{ccc} -1 & 0 & 0 \\ 0 & 1 & 0 \\ 0 & 0 & 1\end{array} \right)
\end{equation}
The $\Gamma$-matrices satisfying the three dimensional Clifford algebra $\{\Gamma_a, \Gamma_b\} =2 \eta_{ab}$ are taken to be:
\begin{equation}
\Gamma_0 =  i \sigma_2 \,, \qquad \Gamma_1 = \sigma_1 \,, \qquad \Gamma_2 = \sigma_3\,,
\end{equation}
where $\sigma_i$ are the usual Pauli matrices:
\begin{equation}
\sigma_1 = \left(\begin{array}{cc} 0 & 1 \\ 1 & 0\end{array} \right)\,, \qquad 
\sigma_2 = \left(\begin{array}{cc} 0 & -i \\ i & 0\end{array} \right)\,, \qquad
\sigma_3 = \left(\begin{array}{cc} 1 & 0 \\ 0 & -1\end{array} \right)\,.
\end{equation}
Finally, the charge conjugation matrix $C = i\sigma_2$, or explicitly
\begin{equation}
C_{\alpha\beta} = \varepsilon_{\alpha\beta} = C^{\alpha\beta}= \left(\begin{array}{cc} 0 & 1 \\ -1 & 0\end{array} \right)\,.
\end{equation}
All spinors in this work are Majorana and the Majorana conjugate of a spinor $\psi^{\alpha}$ is $\bar{\psi}_{\alpha} = C_{\alpha\beta}\psi^{\beta}$. Our conventions imply that we can use the identities
\begin{align}
\Gamma_a\Gamma_b & = \epsilon_{abc}\Gamma^c + \eta_{ab} \mathbb{1}\,, &&& \Gamma^a{}^{\alpha}{}_{\beta} \Gamma_a{}^{\gamma}{}_{\delta} & = 2 \delta^{\alpha}_{\delta} \delta^{\gamma}_{\beta} - \delta^{\alpha}_{\beta}\delta^{\gamma}_{\delta}\,, \\
C^T & = - C\,, &&& C \Gamma_a & = - (\Gamma_a)^T C
\end{align}
In verifying the closure of the supersymmetry algebra on the fields and the off-shell invariance of the action, the three dimensional Fierz relation is useful.
\begin{equation}\label{Fierz}
\zeta\bar{\eta} = - \frac12 \bar{\eta}\, \zeta \, \mathbb{1} - \frac12 (\bar{\eta}\Gamma^a \zeta)\Gamma_a
\end{equation}

\noindent It is sometimes convenient to change basis of the tangent space to one more suited for the $isl(2)$  algebra in the bosonic sector of flat space supergravity. We do this by choosing a map to bring the generators of $SO(2,1)$ ($[J_a,J_b] = \epsilon_{abc}J^c$) to those of $SL(2,\bR)$ satisfying $[L_n,L_m] = (n-m)L_{n+m}$. This defines a matrix $U^a{}_n$ as a map from the tangent space metric $\eta_{ab}$ with $a,b=\{0,1,2\}$ to the metric $\gamma_{nm}$ defined in \eqref{gammadef} with $n,m= \{-1,0,+1\}$, satisfying
\begin{equation}
L_n = J_a\, U^a{}_n \,.
\end{equation}
An explicit representation of $U^a{}_n$ that does the job is for instance
\begin{equation}\label{Umat}
U^a{}_n = \left( \begin{array}{ccc} -1 & 0 & -1 \\ -1 & 0 & 1 \\ 0 & 1 & 0 \end{array} \right)\,.
\end{equation}
In this basis the gamma matrices satisfy a Clifford algebra with
\begin{equation}\label{Cliffnm}
\{\tilde{\Gamma}_m, \tilde{\Gamma}_n\} = 2 \gamma_{nm} \equiv 2 \left( \begin{array}{ccc} 0 & 0 & -2 \\ 0 & 1 & 0 \\ -2 & 0 & 0 \end{array}\right) \qquad \text{with: } n,m= -1 , 0, +1\,.
\end{equation}
A real representation for the gamma matrices with $n,m$ indices can be obtained by taking $\tilde{\Gamma}_n= U^a{}_n \Gamma_a$, or explicitly:
\begin{align}
\tilde{\Gamma}_{-1} & =  - (\sigma_1 + i \sigma_2) = \left(\begin{array}{cc} 0 & -2 \\ 0 & 0\end{array} \right)\,, \\ 
\tilde{\Gamma}_0 & = \sigma_3 = \left(\begin{array}{cc} 1 & 0 \\ 0 & -1\end{array} \right)\,, \\
\tilde{\Gamma}_{+1} & = \sigma_1 - i \sigma_2 = \left(\begin{array}{cc} 0 & 0 \\ 2 & 0\end{array} \right)\,.
\end{align}
In addition to the Clifford algebra \eqref{Cliffnm}, the gamma matrices now satisfy the commutation relations
\begin{equation}\label{Gammacom}
[\tilde{\Gamma}_{n} , \tilde{\Gamma}_{m}] = 2(n-m)\tilde{\Gamma}_{n+m}\,,
\end{equation}
which is just the $sl(2,\bR)$ algebra.


\providecommand{\href}[2]{#2}\begingroup\raggedright\endgroup

\end{document}